\documentclass[aps,prl,twocolumn,reprint,showpacs,amsmath,amssymb,groupaddress,citeautoscript]{revtex4-1}
\usepackage{graphicx}% Include figure files
\begin{document}

%\preprint{APS/123-QED}

\title{Pressure Tuning of the Interplay of Magnetism and Superconductivity  CeCu$_2$Si$_2$}% Force line breaks with \\

\author{E. Lengyel}
 \email{lengyel@cpfs.mpg.de}
\author{M. Nicklas}
 \email{nicklas@cpfs.mpg.de}
\author{H. S. Jeevan}
 \altaffiliation[Present address: ]{I. Physik. Institut, Georg-August-Universit\"{a}t G\"{o}ttingen,
37077 G\"{o}ttingen, Germany.}
\author{C. Geibel}
\author{F. Steglich}

\affiliation{%
Max Planck Institute for Chemical Physics of Solids, N\"{o}thnitzer Str. 40, 01187 Dresden, Germany
}%

\date{\today}% It is always \today, today,

\begin{abstract}

We carried out specific-heat and ac-susceptibility experiments under hydrostatic pressure to
investigate the interplay of spin-density-wave antiferromagnetism ($A$) and superconductivity ($S$)
in single-crystalline $AS$-type CeCu$_2$Si$_2$. We find evidence for a line of magnetic-field- and
pressure-tuned quantum critical points in the normal state in the zero-temperature magnetic field --
pressure plane. Our analysis suggests an extension of this line into the superconducting state and
corroborates the close connection of the underlying mechanisms leading to the formation of the
antiferromagnetic and the superconducting states in $AS$-type CeCu$_2$Si$_2$.

\end{abstract}

% PacS, the Physics and Astronomy
\maketitle

%\section{\label{sec1lev1}Introduction}

% Introduction

The discovery of superconductivity in heavy-fermion (HF) \cite{steg79}, organic \cite{jero80},
cuprate \cite{bedn86}, and most recently in pnictide materials \cite{kami08} changed our
understanding of superconductivity completely. Despite fundamental differences between these
families, the proximity of a magnetic ground-state instability to the superconducting (SC) phase is a
common theme in all of the SC systems. In pnictide, organic, and some HF superconductors itinerant
(spin-density-wave, SDW, type) antiferromagnetism seems to be closely related to the formation of the
SC phase, suggesting a magnetically mediated SC pairing mechanism. The N\'{e}el temperature, $T_N$,
can generally be tuned as function of some external parameter, such as doping or pressure. Typically,
superconductivity develops in the vicinity of the point where $T_N$ disappears as a function of the
external parameter \cite[e.g.][]{jero91,math98,nick04}. The normal state transport and thermodynamic
properties close to that point often disclose a region of non-Fermi-liquid (NFL) behavior hinting at
the presence of a (hidden) quantum critical point (QCP) \cite{math98}. In the cuprates and
iron-pnictides the SC upper-critical-field ($B_{c2}^0$) is generally accessible only by pulsed
magnetic-fields putting strong restrictions on the available experimental probes and their accuracy.
In contrast to these classes of materials, in HF superconductors $B_{c2}^0$ is moderate and the SC
state can be easily suppressed using standard laboratory equipment. This makes the HF materials
perfect model systems for an in-depth investigation of the interplay of magnetism and
superconductivity and a testbed for the theoretical models developed.

CeCu$_2$Si$_2$, the first discovered HF superconductor \cite{steg79} is ideally suited to study the
interplay of SDW antiferromagnetism and superconductivity. The ground state of CeCu$_2$Si$_2$ depends
strongly on the exact stoichiometry. It ranges from (i) antiferromagnetism ($A$-type), coexisting in
a small parameter range with low-$T_c$ superconductivity, via (ii) antiferromagnetism which is
replaced by superconductivity on lowering temperature, or recovered when superconductivity is
suppressed by a sufficiently large magnetic field ($AS$-type), to (iii) solely superconductivity
($S$-type). The antiferromagnetic (AF) order in $A$-type CeCu$_2$Si$_2$ was shown to be an
incommensurate SDW, with a very small ordered moment ($\mu \approx0.1~\mu_{\rm B}$) \cite{stoc04}. In
$AS$-type CeCu$_2$Si$_2$ the AF and SC ordering temperatures are comparable. Resistivity and
specific-heat results obtained in the field-driven low-$T$ normal state of $S$-type CeCu$_2$Si$_2$
revealed NFL phenomena, highly consistent with a three dimensional (3D) SDW QCP \cite{gege98}. The SC
phase in CeCu$_2$Si$_2$ is rather robust against external pressure covering more than 5~GPa.
Introduction of disorder by Ge-doping on the Si-site revealed the presence of two distinct SC domes.
The SC state at low pressures is supposed to be mediated by AF spin fluctuations, while at high
pressures superconductivity is suggested to be mediated by valence fluctuations \cite{yuan03}. An
analysis of thermodynamic data evidences the existence of different SC order parameters in the two
distinct SC phases \cite{leng09}.

In this Letter we will substantiate the close link between superconductivity and antiferromagnetism
in the low-$p$ region in $AS$-type CeCu$_2$Si$_2$. Furthermore, we provide evidence for the existence
of a line of magnetic-field- and pressure-tuned QCP's which extends into the SC region of the
magnetic field -- pressure phase diagram.

Heat-capacity and ac-susceptibility experiments under hydrostatic pressure have been performed on a
single-crystalline sample of $AS$-type CeCu$_2$Si$_2$ ($0.26~{\rm K} \leq T \leq 7$~K, $B_{\parallel
c}\leq8$~T). The resolution of the ac-susceptibility measurements only allows for observing the SC
transition, the AF transition anomaly cannot be resolved. The measurements were carried out in a
single-layer CuBe piston-cylinder type pressure cell (for details see \cite{leng09}).

\begin{figure}[t!]
\includegraphics[angle=0,width=6.8cm,clip]{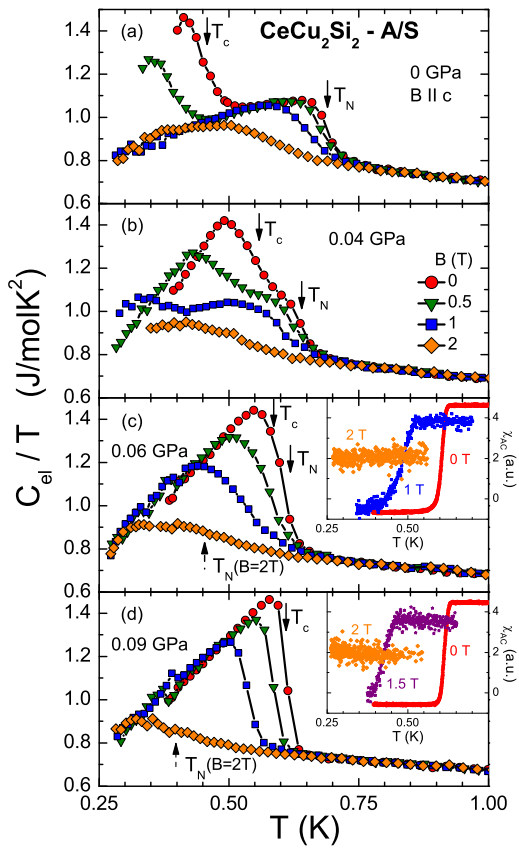}\caption{ \label{Fig1} (Color online) $C_{el}(T)/ T$ as function of
$T$ of $AS$-type CeCu$_2$Si$_2$ for different magnetic fields, $B = 0$~T, 0.5~T, 1~T, and 2~T ($B
\parallel c$) and at four different pressures as indicated in panels a) to d). The solid arrows mark
$T_N$ and $T_c$ at zero magnetic field.  Dashed arrows denote the AF transition at 2~T at 0.06~GPa
and 0.09~GPa. The transition temperatures were determined by an equal-entropy approximation. The
insets of panel c) and d) show $\chi_{ac}(T)$ for different magnetic fields as indicated.}
\end{figure}

Figure~\ref{Fig1} displays the electronic contribution to the specific heat as $C_{el}(T)/T $ at
selected pressures. At atmospheric pressure $AS$-type CeCu$_2$Si$_2$ undergoes two consecutive phase
transitions below $T = 1$~K upon decreasing temperature, the first one at $T_N \approx 0.69$~K, to an
incommensurate AF SDW type of order \cite{stoc04} and the second one at a slightly lower temperature,
marking the onset of superconductivity at $T_c \approx 0.46$~K. The highly enhanced value of the
electronic specific-heat coefficient at low temperatures, $C_{el}/T \approx 0.73$~J/(mol$\rm {K^2}$)
at $T = 0.9$~K, indicates the HF character. A Kondo temperature of $T_K \approx 13$~K can be
determined by analyzing the entropy in the frame of the single-impurity Kondo model \cite{desg82}. At
$T_N$, the entropy reaches a value of only $S_{el}(T_N) \approx 0.11R\ln2$, as anticipated from the
small ordered moment of $\mu_{ord} \approx 0.1~\mu_B$ per Ce atom detected by neutron-diffraction
experiments in the AF state \cite{stoc04}.

Application of a small hydrostatic pressure leads to a rapid shift of $T_N(p)$ to lower temperatures
with an initial slope of ${\rm d}T_N/{\rm d}p\,|_{p=0} \approx -1.17$~K/GPa, while $T_c(p)$ at first
strongly increases (${\rm d}T_c/{\rm d}p\,|_{p=0} \approx 2.33$~K/GPa). In contrast to hydrostatic
pressure, a magnetic field suppresses both $T_N$ and $T_c$, $T_c$ being much more sensitive to the
magnetic field than $T_N$. As a consequence, at 0~GPa and 0.04~GPa the two transition anomalies in
specific heat become more separated in an external magnetic field. At ambient pressure no indication
for a SC transition is present in the data at $B=1$~T anymore in our accessible temperature range
($T>0.27$~K), while at 0.04~GPa the anomaly at $T_c$ is visible in $B=1$~T, but absent in 2~T
indicating an enhanced SC upper-critical field, $B_{c2}^0$, compared with zero pressure. A further
pressure increase to only $p=0.06$~GPa is sufficient to shift the two phase transitions very close to
each other resulting in a single broadened anomaly in $C_{el}(T)/T $. In a closer analysis the
transition temperatures $T_N \approx 0.62$~K and $T_c \approx 0.60$~K can be extracted. Upon
increasing the magnetic field the anomaly in $C_{el}(T)$ further broadens reflecting the different
dependencies of $T_N$ and $T_c$ on the magnetic field. While for $B<2$~T the $\chi_{ac}(T)$ data
confirm the presence of the SC transition, at 2~T no diamagnetic signal is observed anymore proving
that the anomaly observed in specific heat corresponds only to the AF transition and
superconductivity is already suppressed. At a slightly larger pressure, $p = 0.09$~GPa, a single
sharp anomaly in specific heat at $B=0$ signals the transition to the SC state. $\chi_{ac}(T)$
experiments prove the presence of superconductivity up to $B=1.5$~T. We find no hint at a magnetic
transition below $T_c$. At 2~T  $\chi_{ac}(T)$ does not show any diamagnetic signal anymore, but
$C_{el}(T)/T $ exhibits a broad anomaly. Therefore, we identify this anomaly with the transition into
the AF state. At higher pressures (not shown), no indication for a magnetic transition is observed in
the field-induced normal state anymore. Especially, no signature of an AF phase transition inside the
SC state is found at any magnetic field and pressure. Our results clearly indicate that once $T_c(p)$
for a fixed $B$ becomes larger than $T_N(p)$ the presence of the AF phase transition cannot be
detected anymore. This strongly suggests the absence of any long-range magnetic ordering inside the
SC phase. We therefore conclude that the antiferromagnetically ordered state in $AS$-type
CeCu$_2$Si$_2$ is expelled once superconductivity has been established.

\begin{figure}[t!]
\includegraphics[angle=0,width=6.9cm,clip]{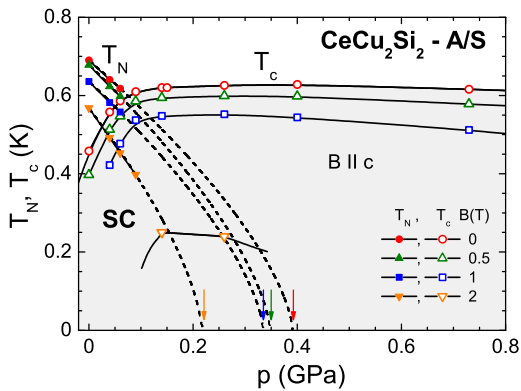}
\caption{\label{Fig2} (Color online) Effect of magnetic field ($B \parallel c$) on the low-pressure
$T - p$ phase diagram of $AS$-type CeCu$_2$Si$_2$. Full symbols correspond to $T_N$, while open
symbols mark $T_c$. The dashed lines extrapolate $T_N(p)$ to zero temperature (see text for details).
The arrows mark the critical pressures $p_c$ for the corresponding magnetic field.}
\end{figure}

The deduced low-pressure $T-p$ phase diagram of $AS$-type CeCu$_2$Si$_2$ is presented in
Fig.~\ref{Fig2}. At zero magnetic field $T_c(p)$ exhibits a weak maximum around
$p_{_{T_{c,max}}}\approx0.4$~GPa. Although, we cannot follow $T_N(p)$ inside the SC state, we can
extrapolate $T_N(p)$ from the normal into the SC state. We follow the predictions of the
spin-fluctuation theory for a 3D SDW QCP and use $T_N(p)\propto(p-p_c)^{2/3}$ to extrapolate $T_N(p)$
\cite{mill93}. By our analysis, we obtain a critical pressure, $p_c \approx 0.39$~GPa, which nearly
coincides with the position of the maximum in $T_c(p)$. We have utilized the same approach to
extrapolate the $T_N(p,B=const.)$ data taken in magnetic fields ($B = 0.5$~T, 1~T, and 2~T). With
increasing magnetic field the critical pressure $p_c(B)$ and the position of the maximum in
$T_c(p,B=const.)$, $p_{_{T_{c,max}}}(B)$, also coincide and shift to lower pressures. The tight
correlation of $p_c(B)$ and $p_{_{T_{c,max}}}(B)$ thus suggests a strong link between the underlying
mechanisms leading to the formation of the two ordered phases in CeCu$_2$Si$_2$.

As pointed out before, we do not detect any magnetic phase-transition anomaly inside the SC state.
Clear evidence for microscopic coexistence of magnetism and (low $T_c$) superconductivity has only
been found at negative chemical pressures \cite{kawa01}. Thus, we speculate that despite the fact
that we cannot identify a magnetic phase transition or a QCP inside the SC state their {\it halo} is
present. For AF spin-fluctuation mediated superconductivity the maximum of the SC dome is expected at
the critical point where magnetism is completely suppressed \cite{naka96,mont99}. This is in
agreement with our findings for $AS$-type CeCu$_2$Si$_2$. However, in zero magnetic field $T_c(p)$
only exhibits a weak maximum at the critical pressure, $p_c\approx0.39$~GPa, in contrast to the
pronounced SC dome typically observed in HF superconductors, e.g. \cite{math98}. The almost pressure
independent behavior of $T_c(p)$ in $AS$-type CeCu$_2$Si$_2$ might be related to the influence of the
additional pairing mechanism provided by the valence fluctuations present at higher pressures
\cite{yuan03,holm04,leng09} and, thus, indicating that already in this relatively low-pressure region
of the phase diagram the two SC pairing mechanisms have to be considered.

\begin{figure}[t!]
\includegraphics[angle=0,width=7cm,clip]{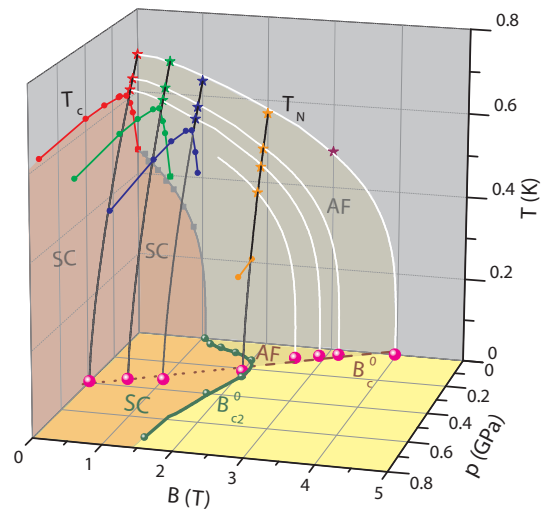}
\caption{\label{Fig3} (Color online) Magnetic and superconducting phase diagram of $AS$-type
CeCu$_2$Si$_2$ as function of $p$ and $B$ for $B\parallel c$. Stars and circles correspond to
$T_N(p,B)$ and $T_c(p,B)$, respectively. Black lines represent fits to the $T_N(p,B=const.)$ data and
white lines to the $T_N(B,p=const.)$ data. A detailed description can be found in the text. The
critical points, $B_c^0(p)$, obtained by the extrapolation of the fitting curves to zero-temperature
are marked with bullets in the $p-B$ plane.  ``SC" and ``AF" mark the superconducting and the
antiferromagnetic phase, respectively. $B_{c2}(T)$ data at ambient pressure is shown. The
extrapolated upper-critical field, $B_{c2}^0(p)$, is indicated by the solid line.}
\end{figure}

Application of both, magnetic field ($B\parallel c$) and pressure, leads to a gradual suppression of
the AF order in $AS$-type CeCu$_2$Si$_2$. While, as already discussed above, for a fixed magnetic
field the $T_N(p)$ curve cannot be followed inside the SC phase anymore, at a constant pressure $T_N$
can be continuously suppressed to zero temperature by increasing the magnetic field suggesting the
presence of field-induced QCP's. The experimental data, $T_N(B,p=const.)$, are well described by the
empirical formula $B_c(T)=B_c^0[1-(T/T_N^0)^n]$, where $T_N^0$ is the N\'{e}el temperature at $B=0$,
$B_c^0=B(T_N=0)$, and $n$ is a fitting parameter. $n$ was determined once at ambient pressure
($n=3.9$) and then kept constant for all other pressures. The results of the fits are indicated by
the white lines in Fig.~\ref{Fig3}. The critical points in the $p-B$ plane at $T=0$, $B_c^0(p)$,
follow a straight line. This is indicated in Fig.~\ref{Fig3} by the dashed line. At ambient pressure
$B_c^0(p=0)\approx 4$~T. An increase of pressure leads to a gradual decrease of $B_c^0(p)$,
$B\parallel c$ (for $B\parallel a$, $B_c^0(p)$ shows initially only a weak pressure dependence
\cite{shei98}). At about $p=0.18$~GPa and $B=2.1$~T the line of critical points hits the SC phase
boundary (solid line in Fig.~\ref{Fig3}), right at the maximum of the upper critical field,
$B_{c2}^0(p)$, suggesting again a close connection of antiferromagnetism and superconductivity in
$AS$-type CeCu$_2$Si$_2$. The estimated critical points inside the SC phase
[$T_N(p,B=const.)\rightarrow 0$, see above], which coincide rather well with the position of the
maximum in $T_c(p)$, lie on the same straight line (dotted line in Fig.~\ref{Fig3}) as the critical
points in the normal state. We note that the HF compound CeRhIn$_5$ exhibits a similar $T$--$p$ phase
diagram at $B=0$ \cite{nick04}. However, in CeRhIn$_5$ the AF state is robust against the application
of a magnetic field, even field-induced magnetism extending into the SC state is observed
\cite{kneb06,park06}. This behavior in magnetic field is in strong contrast to the observations in
CeCu$_2$Si$_2$.

\begin{figure}[t!]
\includegraphics[angle=0,width=6.9cm,clip]{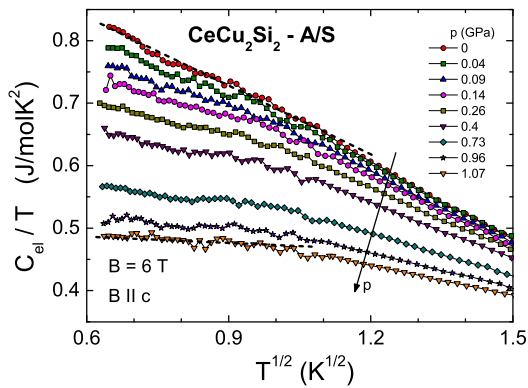}
\caption{\label{Fig4} (Color online) Specific heat of $AS$-type CeCu$_2$Si$_2$ at $B = 6$~T
($B\parallel c$) for different pressures. $C_{el}/T$ is plotted as function of $\sqrt{T}$. The
straight lines correspond to fits of  $C_{el}(T) / T = \gamma_0- a\sqrt{T}$ to the data. For clarity
only two lines for $p=0$ and 1.07 GPa are indicated.}
\end{figure}

To search for further evidence for the presence of a line of field-induced QCP's in the zero
temperature $p-B$ plane we analyzed $C_{el}(T)$ in the normal state. In the proximity of a QCP strong
deviations from Landau-Fermi-liquid (LFL) behavior are expected. Figure~\ref{Fig4} shows the pressure
evolution of $C_{el}/T$ at $B = 6$~T on a $\sqrt{T}$ temperature scale. At ambient pressure,
$C_{el}(T) / T$ increases as $\gamma_0 - a\sqrt{T}$ on lowering the temperature ($T\lesssim1$~K),
showing a clear deviation from $C_{el}(T) / T= const.$ behavior expected for a LFL. This temperature
dependence of $C_{el}(T)$ has been predicted at a 3D AF QCP \cite{mill93, lonz97}. Upon increasing
pressure, the quantum-critical component to the specific heat becomes smaller and smaller. At
sufficiently high pressures ($p \gtrsim 1$~GPa) LFL behavior is observed. The strong deviations from
LFL behavior at low pressures substantiate the proximity to a QCP. As illustrated in Fig.~\ref{Fig3},
upon increasing pressure at a constant $B$ (e.g. $B=6$~T) the distance to the line of QCP's in the
$p-B$ plane continuously increases and, correspondingly, the quantum-critical fluctuations become
weaker until finally LFL behavior is recovered at $p\gtrsim1$~GPa. We want to emphasize the fact that
$T_c(p)$ in zero magnetic field is almost constant ($p>0.1$~GPa), while the quantum-critical
component to the specific heat is strongly reduced upon increasing pressure.

In summary, we studied the magnetic and SC phase diagram of $AS$-type CeCu$_2$Si$_2$ as a function of
external pressure, magnetic field, and temperature. The obtained phase diagram suggests a close
connection of the underlying mechanisms leading to the formation of antiferromagnetism and
superconductivity. Our results indicate the absence of long-range magnetic order in the SC phase and
lead us to the conclusion that antiferromagnetism is expelled once superconductivity is established
in $AS$-type CeCu$_2$Si$_2$. In the normal state $T_N(B,p=const.)$ can be continuously suppressed to
zero temperature by increasing the magnetic field. Hence, we propose the existence of a line of
field-tuned AF QCP's in the zero temperature $p$--$B$ plane. The critical points obtained by
extrapolating $T_N(p,B=const.)$ from the normal into the SC state naturally extend this line.
Furthermore, the positions of the maxima in the pressure dependence of $T_c(p,B=const.)$,
$p_{_{T_{c,max}}}(B)$, coincide with the critical line in accordance with the predictions for AF
spin-fluctuation mediated superconductivity \cite{naka96,mont99}.

We thank G. Sparn who was involved in an early stage of this study. This work was partly supported by
the DFG under the auspices of the Research Unit 960.

\bibliography{bibccs2}% Produces the bibliography via BibTeX.

\end{document}